\documentclass{article} 
\usepackage{indentfirst}
\usepackage{mathrsfs}
\usepackage{amsmath}
\usepackage{booktabs}
\usepackage{makecell}
\usepackage{graphicx}
\usepackage{longtable}
\usepackage{subfigure}
\usepackage{apacite}
\usepackage[font=footnotesize]{caption}
\usepackage{float}
\usepackage[a4paper,left=2.54cm,right=2.54cm,top=2.54cm,bottom=2.54cm]{geometry}
\usepackage{authblk}
\usepackage[T1]{fontenc}
\usepackage[utf8]{inputenc}
\usepackage{abstract}

\newcommand\keywords[1]{\textbf{Keywords}: #1}

\title{On the macroeconomic fundamentals of long-term volatilities and dynamic correlations in COMEX copper futures}
\author[1]{Zian Wang*}
\author[2]{Xinshu Li*}

\affil[1]{Financial Technology Thrust, The Hong Kong University of Science and Technology, Guangzhou, China}
\affil[2]{Department of Mathematics, The Hong Kong University of Science and Technology, Hong Kong, China}

\begin{document}
    \maketitle
\begin{abstract}
    This paper examines the influence of low-frequency macroeconomic variables on the high-frequency returns of copper futures and the long-term correlation with the S\&P 500 index, employing GARCH-MIDAS and DCC-MIDAS modeling frameworks. The estimated results of GARCH-MIDAS show that realized volatility (RV), level of interest rates (IR), industrial production (IP) and producer price index (PPI), volatility of Slope, PPI, consumer sentiment index (CSI), and dollar index (DI) have significant impacts on Copper futures returns, among which PPI is the most efficient macroeconomic variable. From comparison among DCC-GARCH and DCC-MIDAS model, the added MIDAS filter of PPI improves the model fitness and have better performance than RV in effecting the long-run relationship between Copper futures and S\&P 500.
\end{abstract}

\keywords{COMEX copper, GARCH-MIDAS, DCC-GARCH, DCC-MIDAS, macroeconomic variables}

\section{Introduction}
Copper futures are used by miners and dealers to hedge against losses and are in an important position in futures markets all over the world. As an effective conductor of heat and electricity, most copper is used in the electrical sector such as building transmission lines and motors equipment. In the past two decades, the demand of copper has been increasing in developing countries for basic grid construction. As the largest copper importer, China has a huge potential market for electric cars since the production of electric cars requires abundant copper as a raw material. Therefore, a stable and sufficient supply of copper is key to guaranteeing the ongoing shift from gasoline to electric vehicles. Generally, the demand of copper in most developed countries is dominated by manufacturing industry and renewable energy, since it is used to make manufacturing machines and equipment, such as windmills and solar plants. Consequently, copper prices are principally determined by demand of emerging markets such as China and India. 

As copper is associated with many industries, it is often regard as the leading indicator of the world economy. When the global economy is expanding, the copper price tends to rise and vice versa. In recent years, the prices of copper and its derivatives have been fluctuating. While global stock markets have been in recession for the last few years during the pandemic, a soaring rise in copper price took place in 2020. The first reason for the unusual performance is that copper prices are settled in the US dollar, which has fallen sharply during that time. Second, production in major copper exporters such as Chile and Peru have declined due to the COVID-19 pandemic. Thirdly, as having always been the world's largest copper importer for a long time, China did well in the early stages of the pandemic and maintained a sound economic circumstance.

In general, copper has both commodity properties and financial properties, both of which are affected by macroeconomy. Meanwhile, due to the continuous prominence of the financial attributes of the US's commodities, the price volatility of commodity futures market and the price volatility of stock market have increasingly prominent characteristics of synchronization, and the price volatility risk of commodity futures market gradually spreads to the stock market. Therefore, the copper future price is not only affected by the macroeconomic cycle, but also has an increasing correlation with the capital market and stock market. 

Accordingly, it is reasonable using macroeconomic factors as the independent variables of copper future's volatility model. However, most macroeconomic factors have lower frequency than the copper future, which are monthly or quarterly. If we decrease the copper future's frequency, the information contained in the daily variable would be lost. Therefore, this paper uses GARCH-MIDAS (Generalized Auto Regressive Conditional Heteroskedasticity based on Mixed Data Sampling) to study the influence of macroeconomic variables of both monthly and daily frequency to the daily copper future returns' volatility. This method is often used to study the macroeconomic fundamentals of stock market volatility, and has never been previously applied to US copper futures.

In section 5, the univariate GARCH-MIDAS-X models are built with the RV and different macroeconomic factors' (indices) level and volatility. After analysing their independent impact on the copper future returns, bivariate GARCH-MIDAS models take two variables into consideration, and compares the estimation results to those of univariate models to get a general conclusion of the macroeconomic factors' impact on the copper future returns' volatility. In addition to focusing only on copper future itself, this paper further studies the long-run volatility relationship of COMEX copper future and S$\&$P 500. In the context of economic globalization, financial markets are becoming more and more coordinating. Price movements in a single market can spread to another market more easily and quickly. Financial markets are more dependent on each other than before and are affected by fluctuations of other markets. Therefore, this paper uses DCC-MIDAS (Dynamic Conditional Correlation based on Mixed Data Sampling), and applies the efficient macroeconomic factors produced from GARCH-MIDAS model as the MIDAS filter, to estimate the long-run dynamic relationship between copper future and S$\&$P 500 and observe the macroeconomic influence to this relationship.

This paper is organized as follows: In section 2, the literature review on the copper future volatility and the GARCH-MIDAS and DCC-MIDAS is provided. In section 3, we perform the methodologies which describe the GARCH-MIDAS and DCC-MIDAS models. In section 4, we present and process the data of copper future prices, S$\&$P 500 index, two daily and 7 monthly macroeconomic variables. Section 5 covers the empirical analysis. The conclusion follows in section 6.

\section{Literature Review}

\subsection{Copper future and its economic drivers}
Given the importance of copper, a few of studies have already been conducted on the influence factors to the copper's price and its derivatives. Most of existing literatures have applied different models to analyze the variables' effect on copper in three aspects: fundamentals, macroeconomic, and financial variables. Hammoudeh and Yuan (2008) use GARCH family models and reveals the damping effect of interest rate to copper future. Elder (2012) researches on the impact of macroeconomic news on metal futures returns by univariate stepwise linear regression, indicating that the improvement in economic growth has a positive effect on copper returns. The significant influencers he gets include variables of IP (industrial production) and PPI (producer price index). Bunic (2015) applies dynamic model averaging and selection to forecast copper future returns with a large variable set, from which he filters out efficient variables including S$\&$P 500, VIX (volatility index), IP and yield spread. These variables' impact on copper are also proved to be changing over different period. Guzman (2018) compares fundamentals and non-fundamentals variable's influence on copper price by VAR model, and shows the importance of interest rate. Moreover, his result is consistent with Bunic's opinion that the independent variable's impact on copper is explained during a particular period of time. The effectiveness of interest rate is confirmed in the same year that it is an important influence factor to copper price volatility (Duvhammer, 2018). Duvhammer suggest that if the demand of copper and interest rate increase, the volatility of copper would therefore surge. Diaz (2021) researches the economic drivers' effect for predicting copper volatility, and finds that fundamental and financial variables would contribute for prediction, including measures of demand, convenience yields, returns on S$\&$P 500 and some commodities such as oil and gold. He also indicates that the accuracy and effect of the variables is related to time period, indeed, business cycle. More recent researches concern about the period of COVID-19. Zhang (2021) shows the short-term positive effect of the pandemic surge to copper future volatility. Xiao (2022) also reports a correlation between the COVID-19 and copper return, whereas he attributes the increment in copper volatility more to the economic policy.

\subsection{Literatures of GARCH-MIDAS and DCC-MIDAS}

In respective to the mathematical models, GARCH models have won favor of many authors. Kenneth (2003) uses different GARCH models, including standard GARCH, EGARCH, AGARCH and GJR-GARCH, which performs better than previously common random walk model. A research conducted by multivariate 
GARCH models including CCC-AGARCH, VARMA-AGARCH, DCC-AGARCH presents the volatility relationship between stock market and a few commodity markets including copper (Sadorsky, 2014). Sadaorsky also draws the conclusion that the DCC-AGARCH is the best among the three to model the relationship, and to be used for portfolio hedging. They show the importance of the extended models from GARCH, which encourage us to pursue GARCH-MIDAS and DCC-MIDAS. Ghysels et al. (2004) introduced the Mixed Data Sampling regression models (MIDAS), which could apply time series data of different frequencies. MIDAS is applied to a wide range of time series models. The first introduction of MIDAS to GARCH model for stock market return volatility was suggested by Engle et al. (2005). In 2013, they further decompose the long-term component from the stock market return volatility, and use both level and volatility of low frequency macroeconomic variables to build the GARCH-MIDAS model. In their paper, realized volatility is first calculated from the stock market returns with both fixed span and rolling window (Engle et al., 2013). Conrad and Kleen (2019) further dig on the volatility forecasting with GARCH-MIDAS, and compare the performance with a set of GARCH family models. They have a noticeable finding is that as they extend the two covariate mode to three, which shows no better improvement in model fitness. DCC model is introduced by Engle (2002), and is extended to DCC-MIDAS by Colacito (2011).

For the application of GARCH-MIDAS and DCC-MIDAS in the field of copper future returns, Chinese scholars have gained more results. Liu et al. (2019) estimate GARCH-MIDAS models of copper future with interest rate, IP, PPI and M2 respectively, and the theta with interest rate is the greatest and most significant. Li and Mao (2019) examine both SSE A-share market and Shanghai Copper's volatility regarding to four business indices of macro-economic to build univariate GARCH-MIDAS models. DCC-MIDAS is included to further estimate the long-term volatility relationship of the two aimed variables. Wang et al. (2020) apply the GARCH family models to test the efficiency of GEPU in forecasting copper price, and compare the forecast of univariate GARCH-MIDAS model with GEPU and RV to EGARCH and GJR, using DM statistic. With respect to commodity futures volatility in Indian markets, Sreenu et al. (2021) use GARCH-MIDAS to estimate the relationship of Cu future and macroeconomic variables, among which IR (interest rate) and SMI (stock market index) has significant estimated effect on copper future.

There still exists a gap in the U.S. copper future returns modeling with of GARCH-MIDAS model, which we think it has a lot of potential in analyzing the influencers of copper future return volatility with recent 20 years data, since it could include independent variables than the models previously applied.

\section{Methodology}

\subsection{GARCH-MIDAS}
Engle et al. (2013) indicates that the GARCH-MIDAS is based on the spline-GARCH and MIDAS. Apart from that, GARCH-MIDAS uses realized volatility imputing macroeconomic time series to obtain the economic sources of market variance.

In this paper, GARCH-MIDAS decomposes COMEX Copper price volatility into short-term (high-frequency) components and long-term (low-frequency) components. Since macroeconomic expectations have a long-term and profound impact on the futures market, monthly macroeconomic variables can be used to construct the long-term part of the model; the short-term component is mainly affected by short-term factors such as the daily liquidity of the futures market and current trading information that follows the GARCH (1,1) process.

The log return on day i in month t is written as follows with the conditional expectation,
\begin{equation}
    \mu = E_{i-1,t}(r_{i,t})
\end{equation}
\begin{equation}
    r_{i,t} = \mu + \sqrt[]{\tau _t\cdot g_{i,t}}\varepsilon _{i,t},\forall i = 1,2,\dots ,N_t
\end{equation}
where $g_{i,t}$ is the short-term high frequency component of the conditional variance of $r_{i,t}$, and the conditional variance of $r_{i,t}$ is the product of the long-term component $\tau_t$ and the short-term component $g_{i,t}$. $\Phi _{i-1,t}\sim N(0,1)$ with $\Phi _{i-1,t}$ is the information set in the day i-1 of period $t$.

For the short-term component, Engle et al. (2013) defines that it is followed by GARCH (1,1), while Conrad and Kleen (2019) assume it to follow a mean-reverting unit-variance GJR-GARCH (1,1) as below.

\begin{equation}
    g_{i,t} = \left ( 1-\alpha -\frac{\gamma }{2}-\beta \right ) +\left ( \alpha +\gamma _{\left \{ \varepsilon _{i-1}<0 \right \} } \right ) \frac{\varepsilon _{i-1,t}^2}{\tau _t} +\beta \cdot g_{i-1,t}
\end{equation}

For the long-term component based on MIDAS, it is specified by the realized volatility $RV$,

\begin{equation}
    \tau = m+\theta \sum_{K}^{k = 1} \phi _k(\omega_1,\omega _2)RV_{t-k}
\end{equation}
\begin{equation}
    RV_t = \sum_{i = 1}^{N_t} r^2_{i,t}
\end{equation}
where m is a constant term, k represents the number of lags, and K is the maximum value of the realized volatility. $ \theta $ represents the impact of the $RV_t$ lag term to the long-term component $\tau_t$.

For the weighting scheme, we have:
\begin{equation}
    \phi_k(\omega )=\left\{
    \begin{aligned}
         & \frac{(k-K)^{\omega _1-1}(1-\frac{k}K)^{\omega _2-1}}{\sum_{j=1}^{K}(\frac{j}K^{\omega _1-1})(1-\frac{j}K)^{\omega _2-1} }Beta \\
         & \frac{\omega ^k}{\sum_{j=1}^{K}\omega ^j } Exp.weighted
    \end{aligned}
    \right.
\end{equation}
and generally, we set $\omega_1$ equal to 1. To ensure that the weight of the lag variable is in the form of attenuation (the closer to the current period, the greater the impact on the current period), the coefficient $\omega_2$ determines the attenuation rate of the influence degree of low frequency data on high frequency data.

Similarly, the long-term component $\tau _t$ based on single factor is easily expanded to the model based on multiple macroeconomics variables,

\begin{equation}
    \log\tau_t=m+\sum_{j=1}^{M} \theta _j\sum_{k=1}^{K} \phi _k(\omega _{1,j},\omega _{2,j})X^j_{t-k}
\end{equation}
where $M$ is the number of macroeconomics determinants, $X_{t-k}^j$ refers to the volatility of macroeconomic variables lagging $K$ period relative to the current period $t$. The logarithm guarantees that the long-term component of the macroeconomic variable is always positive.

The likelihood function that estimates multiple macroeconomics variables is

\begin{equation}
    LLF=-\frac{1}{2} \left [ (2\pi)^{TN}+\sum_{t=1}^{T}\sum_{i=1}^{N}\ln (g_{i,t}\tau_t)+\sum_{t=1}^{T}\sum_{i=1}^{N}\frac{(r_{i,t}-\mu )^2}{g_{i,t}\tau _t}      \right ]
\end{equation}
where $T$ is the number of months, and $N$ refers to the days of each month.

\subsection{DCC-GARCH}

Engle (2001) introduced the dynamic conditional correlation (DCC) that is regarded as an improved abstraction of the constant conditional correlation (CCC). The constant conditional correlation defines
\begin{equation}
    H_t=D_tRD_t
\end{equation}
\begin{equation}
    D_t=diag\left \{ \sqrt[]{h_{i,t}}  \right \}
\end{equation}
\begin{equation}
    h_{i,t}=E_{t-1}(r^2_{i_1t})
\end{equation}
where $R$ is the conditional correlation matrix. The DCC estimator requires the conditional correlation matrix $R$ to be varied as time goes by
\begin{equation}
    H_t=D_tR_tD_t
\end{equation}
\begin{equation}
    R_t=diag(Q_t)^{-\frac12}Q_tdiag(Q_t)^{-\frac12}
\end{equation}
\begin{equation}
    \rho _{i,q,t}=\frac{q_{i,q,t}}{\sqrt[]{q_{ii,t}q_{jj,t}} }
\end{equation}
\begin{equation}
    Q_t=S(1-\alpha -\beta )+\alpha (\varepsilon _{t-1}\varepsilon _{t-1}^\prime)+\beta Q_{t-1}
\end{equation}
where $\varepsilon _{t-1}^\prime=diag(Q_t)^{-\frac12}\varepsilon _{t-1}$, and $\varepsilon _{t-1}$ is the standardized residual obtained by establishing univariate GARCH (1,1) model. The covariance matrix $Q_t$ with $q_{i,j,t}$ is the exponentially weighted moving average process of $\varepsilon _{t-1}^\prime$. Hence, the model is mean reverting if $\alpha+\beta<1$, and the unconditional correlation matrix of $\varepsilon$ is as below.

\begin{equation}
    \begin{aligned}
        S=E(\varepsilon_t\varepsilon_{t-1}^\prime) \\
    \end{aligned}
\end{equation}

Therefore, we primarily estimate the returns of time series by univariate GARCH (1,1) model to get the volatility. Then we obtained the time varying volatilities of two different returns and the dynamic conditional correlation between them by DCC-GARCH model.

\subsection{DCC-MIDAS}

Dynamic correlations are a natural extension of the GARCH-MIDAS model to the Engle (2002) DCC model. In DCC-MIDAS, there still exists one long-run and one short-run component—name $q_{i,j,t}$ as the short-run correlation between assets $i$ and $j$, whereas $\bar{\rho}_{i,j,t}$ is a slowly moving long run correlation. In our case, i and j refer to Cu future returns and S\&P 500 return. Since we have only two assets to be involved in the correlation model, the general form of model written in matrix is not presented.

\begin{equation}
    q_{i,j,t}=\bar{\rho }_{i,j,t}(1-a-b)+a\xi _{i,t-1}\xi_{j,t-1}+bq_{i,j,t-1}
\end{equation}
\begin{equation}
    \bar{\rho }_{i,j,t}=\sum_{l=1}^{k_c^{ij}} \varphi _l(\omega _r^{ij})c_{i,j,t-1}
\end{equation}
\begin{equation}
    c_{i,j,t}=\frac{ \sum\limits _{k=t-N_c^{ij} }^{t}\xi_{i,k}\xi_{j,k} }{\sqrt[]{\sum\limits _{k=t-N^c_{ij}}^{t}\xi^2_{i,k} }\sqrt[]{\sum\limits _{k=t-N^c_{ij}}^{t}\xi^2_{j,k} } }
\end{equation}

The weighting scheme is similar to $\phi_k(\omega)$ in equation (6), just using a different subscript to differ. $\xi_{i,t}$ are the standardized residuals, and their simple products are used to formulate $c_{i,j,t}$. $N_c^{ij}$ is the number of lag lengths of the standardized residuals $\xi_{i,t}$ to form the long-run correlation, while $K_c^{ij}$ is the span lengths of historical correlations. The a and b are parameters to be estimated, so as the weighting parameter $\omega$.

\subsection{Variance Ratio}
The variance ratio is defined as the fraction of the sample variance of the log of total quarterly conditional volatility, which can be explained by the sample variance of the log long-term component (Engle et al., 2013). For the GARCH-MIDAS models, we have
\begin{equation}
    VR(X)=\frac{Var\left ( \log(\tau _t^{\left [ M \right ] }) \right ) }{Var\left ( \log(\tau _t^{\left [ M \right ] }g _t^{\left [ M \right ] }) \right ) }
\end{equation}

\section{Data}

This paper collects daily COMEX copper prices, several stock indexes, and monthly macroeconomic fundamentals, which measure different aspects of the U.S. economy concerning inflation, interest, currency, etc. Although some monthly macroeconomic ones have some missing value, they are all from Sep 3, 2002 to Aug 31, 2022. The data in this paper are collected from the Wind database.

COMEX copper price is the continuous daily price of copper future. PPI is the one by all commodities without seasonal adjustment. The slope is the yield spread between 10-year treasury bond and 3-month treasury bill. IR is the rate of return about 3-month treasury bill. DI is the nominal dollar index. IP is the one by all industries without seasonal adjustment. PMI is the manufactural one taken from the Institute for Supply Management (ISM). CSI is the one from Michigan University. NOI is from ISM as well. NAI is the Chicago Fed national activity index.

Guzmán and Silva (2018) state that the copper price is significantly affected by 3 macroeconomic factors: liquidity in the USA, interest rates, and the dollar index. The data in Panel B of Table 1 also contains the influence in the present and prospective markets respectively.

\begin{table}[H]
    \footnotesize
    \centering
    \caption{Data Processing and Stationarity}
    \label{Table1}
    \resizebox{\textwidth}{!}{%
        \begin{tabular}{llllll}
            \hline
            \textbf{Variable}                                                                            & \textbf{Code in Model} & \textbf{Frequency} & \makecell[l]{\textbf{Pre-processing}                              \\\textbf{Method}} & \multicolumn{2}{c}{\textbf{ADF Test}} \\ \hline
            \multicolumn{4}{l}{\textbf{Panel A: Daily COMEX Copper Price Data(Sep 3, 2002-Aug 31, 2022)}} & ADF                    & P value                                                                                \\
            COMEX Copper Price                                                                           & Cu                     & Daily              & log-differencing                     & -20.7322(-3.4317) & 0.0000 \\
            Standard \& Poor's 500 Index                                                                 & S\&P 500               & Daily              & log-differencing                     & -17.4079(-3.4317) & 0.0000 \\ \hline
            \multicolumn{4}{l}{\textbf{Panel B: Macro data(Sep 3, 2002-Aug 31, 2022)}}                                           &                        &                                                                                        \\
            \multicolumn{4}{l}{\textbf{Inflation\&Interest-rate Variables}}                              &                        &                                                                                        \\
            Producer Price Index                                                                         & PPI                    & Monthly            & log-differencing                     & -7.1125(-3.4582)  & 0.0000 \\
            Term Spread                                                                                  & Slope                  & Daily              & First-differencing                   & -14.5073(-3.4317) & 0.0000 \\
            Short-term Interest Rate                                                                     & IR                     & Daily              & First-differencing                   & -9.4308(-3.4317)  & 0.0000 \\
            \multicolumn{4}{l}{\textbf{Currency}}                                                        &                        &                                                                                        \\
            Dollar Index                                                                                & DI                     & Monthly            & log-differencing                     & -6.0197(-3.4589)  & 0.0000 \\
            \multicolumn{4}{l}{\textbf{Current-stance-of-the-economy Variables}}                         &                        &                                                                                        \\
            Industrial Production Index                                                                  & IP                     & Monthly            & log-differencing                     & -3.6624(-3.4599)  & 0.0047 \\
            Purchasing Managers Index                                                                    & PMI                    & Monthly            & log-differencing                     & -13.9861(-3.4581) & 0.0000 \\
            Consumer Sentiment Index                                                                     & CSI                    & Monthly            & log-differencing                     & -13.8628(-3.4582) & 0.0000 \\
            \multicolumn{4}{l}{\textbf{Future-stance-of-the-economy Variables}}                          &                        &                                                                                        \\
            The New Orders Index                                                                         & NOI                    & Monthly            & log-differencing                     & -5.3468(-3.4596)  & 0.0000 \\
            National Activity Index                                                                      & NAI                    & Monthly            & None                                 & -11.8692(-3.4581) & 0.0000 \\ \hline
        \end{tabular}%
    }
\end{table}
\noindent
\textit{Note: The number in parenthesis represents lower critical value of the  1$\%$ significance level.}

To make the macroeconomic variables stationary before being inputted, it is necessary to do some pre-processing based on the data type of each one. Especially, the national activity index does not need any adjustment, since it is already stationary.

Figure 1 illustrates that the return of copper futures volatile a little stronger than the other in Panel A. The standard deviation of the COMEX copper returns is 0.0174, slightly higher than the 0.0122 of S\&P 500 returns. According to the Ljung-Box test, they are not white noise. The static Spearman correlation coefficient of COMEX copper returns with S\&P 500 returns is 0.27. And the dynamic conditional correlation will be analyzed in the DCC-GARCH part.

\begin{figure}[H]
    \centering
    \caption{Returns on Copper and the S$\&$P 500}
    \includegraphics[width=\textwidth]{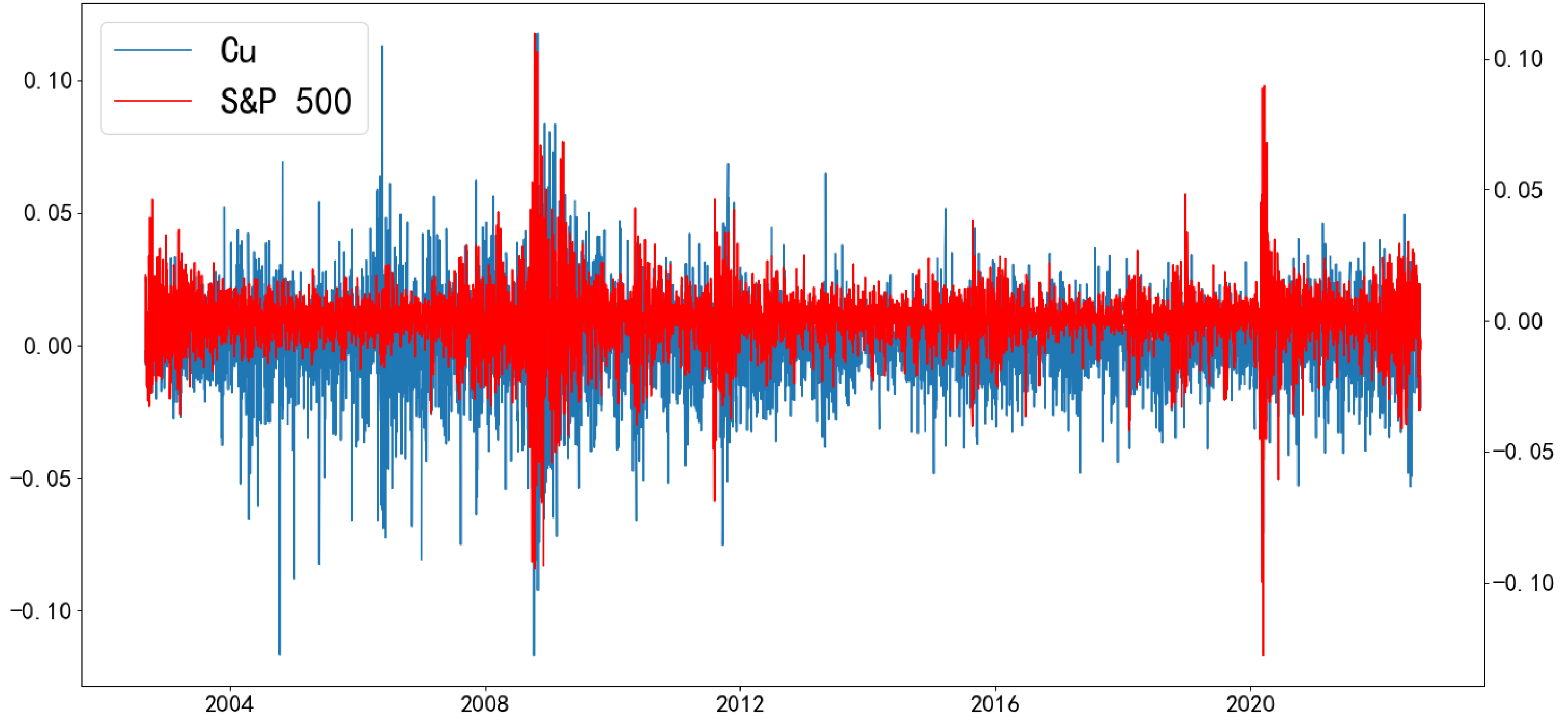}
\end{figure}

Table \ref{Table2} provides basic descriptive statistics about all factors after pre-processing in terms of observations, minimum value, maximum value, standard deviation, skewness, and kurtosis. Simultaneously, we choose the Kolmogorov-Smirnov test and Jarque-Bera test to test normality. The result is that all variables are not normally distributed. Apart from that, no time series is white noise by the Ljung-Box test lags with 20, and they all have an ARCH effect.

\begin{table}[H]
    \footnotesize
    \centering
    \caption{Descriptive Statistics for Copper Futures and Macro Data}
    \label{Table2}
    \resizebox{\textwidth}{!}{%
        \begin{tabular}{cllllllllll}
            \hline
            \multicolumn{1}{l}{} & \multicolumn{1}{c}{\textbf{Obs.}} & \multicolumn{1}{c}{\textbf{Min}} & \multicolumn{1}{c}{\textbf{Max}} & \multicolumn{1}{c}{\textbf{Mean}} & \multicolumn{1}{c}{\textbf{SD}} & \multicolumn{1}{c}{\textbf{Skew.}} & \multicolumn{1}{c}{\textbf{Kurt.}} & \multicolumn{1}{c}{\makecell[l]{\textbf{K-S}                   \\\textbf{pvalue}}} & \multicolumn{1}{c}{\makecell[l]{\textbf{J-B}\\\textbf{pvalue}}} & \multicolumn{1}{c}{\makecell[l]{\textbf{L-B}\\\textbf{pvalue}}} \\ \hline
            \multicolumn{11}{l}{\textbf{Panel A}}                                                                                                                                                                                                                                                                                                           \\
            \textbf{Cu}          & 4990                              & -0.1169                          & 0.1177                           & 0.0003                            & 0.0174                          & -0.2195                            & 3.9983                             & 0.0000                                       & 0.0000 & 0.0000 \\
            \textbf{S\&P 500}    & 5034                              & -0.1277                          & 0.1096                           & 0.0003                            & 0.0122                          & -0.5014                            & 12.5807                            & 0.0000                                       & 0.0000 & 0.0000 \\
            \multicolumn{11}{l}{\textbf{Panel B}}                                                                                                                                                                                                                                                                                                           \\
            \textbf{IP}          & 239                               & -0.1683                          & 0.0881                           & 0.0006                            & 0.0210                          & -1.8276                            & 17.6488                            & 0.0000                                       & 0.0000 & 0.0029 \\
            \textbf{PPI}         & 239                               & -0.0548                          & 0.0316                           & 0.0030                            & 0.0128                          & -0.9600                            & 3.2381                             & 0.0000                                       & 0.0000 & 0.0000 \\
            \textbf{IR}          & 5000                              & -0.8100                          & 0.7600                           & 0.0003                            & 0.0450                          & -0.8040                            & 79.6801                            & 0.0000                                       & 0.0000 & 0.0000 \\
            \textbf{Slope}       & 5000                              & -0.5200                          & 0.7400                           & -0.0004                           & 0.0651                          & 0.3052                             & 12.1365                            & 0.0000                                       & 0.0000 & 0.0000 \\
            \textbf{PMI}         & 239                               & -0.1682                          & 0.1992                           & 0.0002                            & 0.0387                          & -0.0126                            & 4.3484                             & 0.0000                                       & 0.0000 & 0.1399 \\
            \textbf{CSI}         & 239                               & -0.2159                          & 0.1276                           & -0.0016                           & 0.0570                          & -0.5481                            & 1.1135                             & 0.0000                                       & 0.0000 & 0.9188 \\
            \textbf{NOI}         & 239                               & -0.4429                          & 0.5730                           & -0.0001                           & 0.0860                          & 0.6719                             & 10.9757                            & 0.0000                                       & 0.0000 & 0.4811 \\
            \textbf{NAI}         & 240                               & -17.9600                         & 6.1200                           & -0.1145                           & 1.4135                          & -8.1356                            & 108.7495                           & 0.0000                                       & 0.0000 & 0.0178 \\
            \textbf{DI}          & 239                               & -0.0337                          & 0.0628                           & 0.0003                            & 0.0127                          & 0.5522                             & 2.1008                             & 0.0000                                       & 0.0000 & 0.0000 \\ \hline
        \end{tabular}%
    }
\end{table}
\noindent
\textit{Note: The data in Panel A are calculated to return in this table, and the macroeconomic variables are stationary after processing. For the K-S test and J-B test, if the p-value is bigger than 0.05, the time series will be regarded as normally distributed. For the L-B test, the data is not going to be white noise if the p-value is smaller than 0.05.}

\section{Empirical Analysis}

In this section, we apply the GARCH-MIDAS model to the copper future's daily return, based on its monthly realized volatility (RV), seven monthly macroeconomic variables and two daily macroeconomic variables respectively. Then based on the results, we apply the DCC-MIDAS model to copper future's daily return and S\&P 500 return, with a significant macroeconomic variable filtered by GARCH-MIDAS.

\subsection{GARCH-MIDAS-RV Model with Realized Volatility }

Table \ref{Table3} shows the estimated parameters of GARCH-MIDAS model with monthly realized volatility of copper future’s return in the first row. Following Conrad (2015), by comparing the log likelihood of GARCH-MIDAS of monthly RV with lag length K from 12 to 60 (one to 5 MIDAS years), the lag length K of long-term component we apply is 12 with the greatest log likelihood value, while the estimation results performed robust for larger lag lengths.
\begin{table}[H]
    \footnotesize
    \centering
    \caption{Estimated results of realized volatility (RV)}
    \label{Table3}
    \resizebox{\textwidth}{!}{%
        \begin{tabular}{@{}llllllllll@{}}
            \toprule
               & $\mu$     & $\alpha$  & $\beta$   & m         & $\theta$  & $\omega_2$ & BIC      & LLH      & VR(X) \\ \midrule
            RV & 0.0265    & 0.0790*** & 0.8598*** & 0.5167*** & 0.0076*** & 1.3866**   & 17724.07 & -8832.45 & 44.99 \\
               & (-0.2125) & (-0.0038) & (0.0000)  & (0.0000)  & (0.0000)  & (-0.0241)  &          &          &       \\ \bottomrule
        \end{tabular}%
    }
\end{table}
\noindent
\textit{Note: Numbers in parenthesis are the p-value of each estimated parameter. ***, **, and * indicate significance at the 1$\%$, 5$\%$, and 10$\%$ levels, respectively. VR(X) is variance ratio test. LLH is the result of maximized log likelihood.}

Both p-value of alpha and beta are extremely close to 0, which shows their significance of existence. Besides, the sum of $\alpha$ and $\beta$ is smaller than and close to 1, indicating the volatility persistence in short-run GARCH term. Also, the parameter $\theta$ is noticeably significant and is positive, which shows that the monthly realized volatility has positive influence to the volatility of copper future's return. Following Engle (2013) and Conrad (2015), we use the restricted weighting scheme with $\omega_1$=1. Another reason to restrict $\omega_1$ is that the bivariate model only allows one independent variable's weighting scheme to be unrestricted, so we uniformly keep both variables restricted, also for the univariate model.

Figure 2 shows the estimated conditional volatility and its long-run component of the GARCH-MIDAS model with monthly fixed RV and 12 lagged monthly RVs in the MIDAS filter. The long run component matches the dynamic of conditional volatility very well.

\begin{figure}[H]
    \centering
    \caption{Conditional volatility and its long run component of Cu future returns-Based on RV}
    \includegraphics[width=\textwidth,height=0.4\textheight]{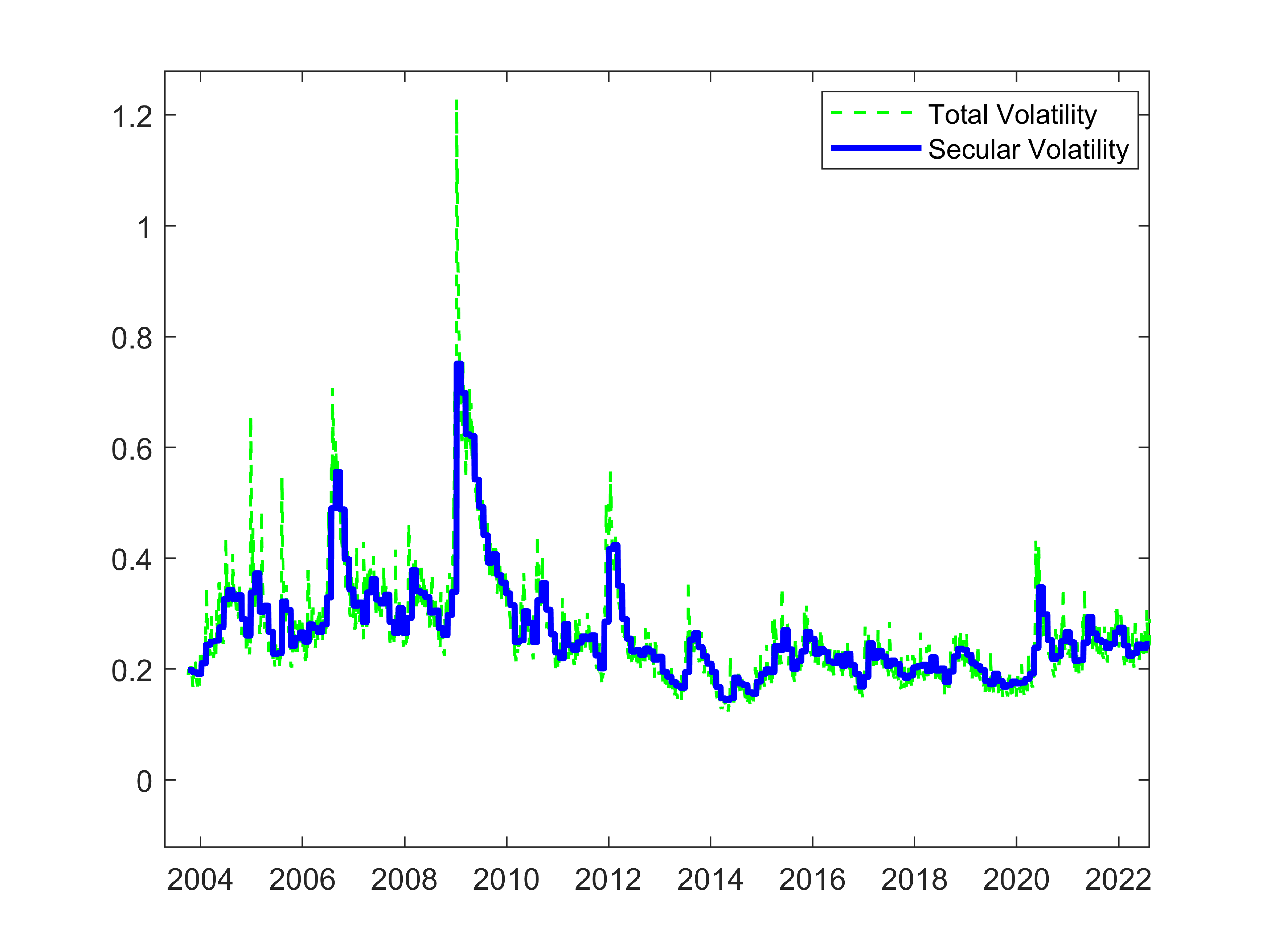}
\end{figure}

\subsection{GARCH-MIDAS-X Model with Macroeconomic Variables}

For the macroeconomic variables, we consider both their level and volatility's effect to the copper future return. The level values of all monthly macroeconomic variables except NAI are processed by log differencing, similar to the log return calculation of copper future prices, while NAI is already stationary without differencing and has negative values which cannot be logged. For the daily variables, IR and Slope, which are not indices, become stationary after first differencing, as illustrated in Table \ref{Table1}. The ADF test results of (log) differenced macroeconomic variables are shown in Table \ref{Table1}. For volatilities, we fit $AR(p): Y_t=\varphi_1Y_{t-1}+\varphi_2Y_{t-2}+\dots+\varphi_pY_{t-p}+\varepsilon_t$ to the original index series and take the square of residuals as simulated macroeconomic factor's volatility $\hat{\sigma}_t^2=\varepsilon_t^2$, following Engle's (2013) approach.

Firstly, univariate GARCH-MIDAS model is applied to estimate each macro\-economic parameter. The data have been adjusted to the intersection of the trading days of daily variables and copper futures, with less than 50 days to be deleted. For the monthly data of low frequency, we keep their value the same for the whole month according to the business days of copper future so that they have same size and aligned by month. To keep consistence with RV, the lag length K of monthly variables is 12. Therefore, the K for daily variables is 250, suggesting 250 trading days in a year.

We have computed the estimation results with GARCH-MIDAS model of RV, seven monthly variables and two daily variables respectively. Their $\alpha$ and $\beta$ are all significant and have sums over 0.98 close to 1, which shows their noticeable GARCH characteristic. For all the independent variables, their conditional mean  (or intercept term) are insignificant and m terms (constant term in long run component ) are significant. Therefore, the estimation of  terms are left out here and in the following tables as they are all insignificant. The  term indicates the effect of external macroeconomic variables' variation to the copper future return. As known, BIC and log likelihood measures the fitness of the model. VR(X) is the variance ratio (Engle et al., 2013), which presents to which extent the variation in the squared returns can be attributed to the variation in the long-term component. The estimation results of univariate model with the returns and volatility of macroeconomic variables are presented in Table \ref{Table4}.
{
    \begin{longtable}{lllllllll}
    \caption{Univariate GARCH-MIDAS model}
    \label{Table4}\\
    \toprule
                                   & $\alpha$  & $\beta$   & m         & $\theta$  & $\omega_2$ & BIC      & LLH      & VR(X) \\ \midrule
    RV                             & 0.0790*** & 0.8598*** & 0.5167*** & 0.0076*** & 1.3866**   & 17724.07 & -8832.45 & 44.99 \\
                                   & (0.0038)  & (0.0000)  & (0.0000)  & (0.0000)  & (0.0241)   &          &          &       \\ \midrule
    \multicolumn{9}{l}{Index Level (log difference)}                                                                          \\
    IR                             & 0.0328*   & 0.9637*** & -0.5335   & -0.2759** & 8.6831***  & 17736.24 & -8838.53 & 8.81  \\
                                   & (0.0784)  & (0.0000)  & (0.9098)  & (0.0260)  & (0.0081)   &          &          &       \\
    Slope                          & 0.0514*** & 0.9340*** & 1.0198*** & -0.0427   & 1.2481     & 17737.36 & -8839.09 & 0.09  \\
                                   & (0.0060)  & (0.0000)  & (0.0000)  & (0.9398)  & (0.6180)   &          &          &       \\
    IP                             & 0.0494*** & 0.9367*** & 1.0098*** & -0.0273*  & 70.5264*** & 17739.03 & -8839.92 & 0.99  \\
                                   & (0.0022)  & (0.0000)  & (0.0000)  & (0.0929)  & (0.0001)   &          &          &       \\
    PPI                            & 0.0514*** & 0.9302*** & 0.8991*** & 0.3340**  & 1.8525*    & 17734.79 & -8837.81 & 14.46 \\
                                   & (0.0017)  & (0.0000)  & (0.0000)  & (0.0147)  & (0.0539)   &          &          &       \\
    PMI                            & 0.0509*** & 0.9347*** & 1.0158*** & -0.0101   & 3.1346     & 17744.13 & -8842.48 & 0.09  \\
                                   & (0.0035)  & (0.0000)  & (0.0000)  & (0.9163)  & (0.9229)   &          &          &       \\
    CSI                            & 0.0514*** & 0.9338*** & 1.0167*** & 0.0061    & 3.5128**   & 17744.11 & -8842.47 & 0.04  \\
                                   & (0.0021)  & (0.0000)  & (0.0000)  & (0.7975)  & (0.0114)   &          &          &       \\
    NOI                            & 0.0511*** & 0.9345*** & 1.0100*** & -0.0095   & 4.7475     & 17743.34 & -8842.08 & 0.38  \\
                                   & (0.0030)  & (0.0000)  & (0.0000)  & (0.5948)  & (0.9122)   &          &          &       \\
    NAI                            & 0.0492*** & 0.9360*** & 1.0258*** & 0.0024    & 1.4227     & 17741.85 & -8841.33 & 4.17  \\
                                   & (0.0029)  & (0.0000)  & (0.0000)  & (0.2324)  & (0.1304)   &          &          &       \\
    DI                             & 0.0528*** & 0.9304*** & 1.0175*** & -0.1435   & 1.5363*    & 17743.25 & -8842.04 & 1.62  \\
                                   & (0.0023)  & (0.0000)  & (0.0000)  & (0.5339)  & (0.0752)   &          &          &       \\ \midrule
    \multicolumn{2}{l}{Volatility} &           &           &           &           &            &          &                  \\
    IR                             & 0.0292    & 0.9678*** & -0.5066   & 0.3805    & 3.9167     & 17752.87 & -8846.85 & 9.38  \\
                                   & (0.7066)  & (0.0000)  & (0.9791)  & (0.1503)  & (0.1538)   &          &          &       \\
    Slope                          & 0.0564*** & 0.9209*** & 0.6866*** & 0.7715*** & 1.2858     & 17724.89 & -8832.86 & 31.28 \\
                                   & (0.0025)  & (0.0000)  & (0.0000)  & (0.0000)  & (0.1761)   &          &          &       \\
    IP                             & 0.0513*** & 0.9336*** & 1.0109*** & 0.0000    & 1.0003     & 17744.20 & -8842.51 & 0.01  \\
                                   & (0.0022)  & (0.0000)  & (0.0000)  & (0.9080)  & (0.7632)   &          &          &       \\
    PPI                            & 0.0861*** & 0.8698*** & 0.7793*** & 0.0006**  & 1.0000     & 17750.60 & -8845.71 & 26.38 \\
                                   & (0.0077)  & (0.0000)  & (0.0000)  & (0.0308)  & (0.1120)   &          &          &       \\
    PMI                            & 0.0505*** & 0.9332*** & 1.0370*** & 0.0000    & 16.1558    & 17742.36 & -8841.59 & 0.69  \\
                                   & (0.0025)  & (0.0000)  & (0.0000)  & (0.2582)  & (0.3510)   &          &          &       \\
    CSI                            & 0.0574*** & 0.9058*** & 0.6596*** & 0.0001*   & 1.0000***  & 17743.88 & -8842.35 & 11.15 \\
                                   & (0.0073)  & (0.0000)  & (0.0011)  & (0.0903)  & (0.0024)   &          &          &       \\
    NOI                            & 0.0401*** & 0.9415*** & 0.9823*** & 0.0000    & 9.5124     & 17745.03 & -8842.92 & 0.78  \\
                                   & (0.0016)  & (0.0000)  & (0.0000)  & (0.3214)  & (0.4872)   &          &          &       \\
    NAI                            & 0.0493*** & 0.9366*** & 1.0197*** & 0.0000    & 5.0292*    & 17741.98 & -8841.40 & 1.33  \\
                                   & (0.0036)  & (0.0000)  & (0.0000)  & (0.2971)  & (0.0886)   &          &          &       \\
    DI                             & 0.0754**  & 0.8924*** & 0.7098*** & 0.0028*   & 1.0000     & 17742.54 & -8841.68 & 15.28 \\
                                   & (0.0101)  & (0.0000)  & (0.0017)  & (0.0585)  & (0.4003)   &          &          &       \\ \bottomrule
\end{longtable}
}
\noindent
\textit{Note: Numbers in parenthesis are the p-value of each estimated parameter. ***, **, and * indicate significance at the 1$\%$, 5$\%$, and 10$\%$ levels, respectively. Estimates for $\mu$ and $\gamma$ are omitted.}

For the index levels (after (log) differencing), $\theta$s of IR and PPI are significant on 0.05 level, and that of IP is significant at 0.1 level. The $\theta$ of IR and IP are negative, indicating the increment in their value would decrease the volatility of copper future return, while the positive $\theta$ of PPI shows the opposite. As illustrated by Schwert (1989), the returns of IP shows increase in during recessions, which is a countercyclical pattern. Moreover, as PPI is often considered as inflation indicator, its positive $\theta$ has captured the increasing inflation of the observed period. The criterion values of the three significant variables are also close. However, their variance ratios are quite different, while PPI return's VR of 14.46 is the largest among all macroeconomic variables, showing its importance. The weighting scheme $\omega_2$ of IP is noticeably large with value of 70.5264, under the restricted weighting scheme ($\omega_1$=1).

The estimations of models with respect to volatilities are shown in the lower section of Table \ref{Table4}. PPI is the only one with significant $\theta$ of both its level and volatility, but its $\theta$ of volatility is much smaller and has less positive effect on copper future's return. Its $\omega_2$ is close to 1 but insignificant, while in the unrestricted weighting scheme, PPI's $\omega_1$=1.75 and $\omega_2$=1.02, both significant. From this we infer that PPI's volatility should fit better to the unrestricted weighting scheme, which is a limitation of provided package that it cannot release the weighting scheme of bivariate model. The estimated $\omega_2$ of CSI's volatility is significantly close to 1, under the restricted weighting scheme. Although the four variables' $\theta$ are all significant, only Slope's $\theta$ deviates from zero and have the value 0.7715, indicating positive influence to the copper future return's volatility. Moreover, the GARCH-MDIAS-X model with Slope's volatility shows good fitness with VR of 31.28, higher than that of PPI volatility's 26.38.

\begin{figure}[H]
    \captionsetup{font=footnotesize}
    \caption{Conditional volatility and its long run component of Cu future returns - Based on Index Returns}
    \label{Figure3}
    \includegraphics[width=\textwidth,height=0.5\textheight]{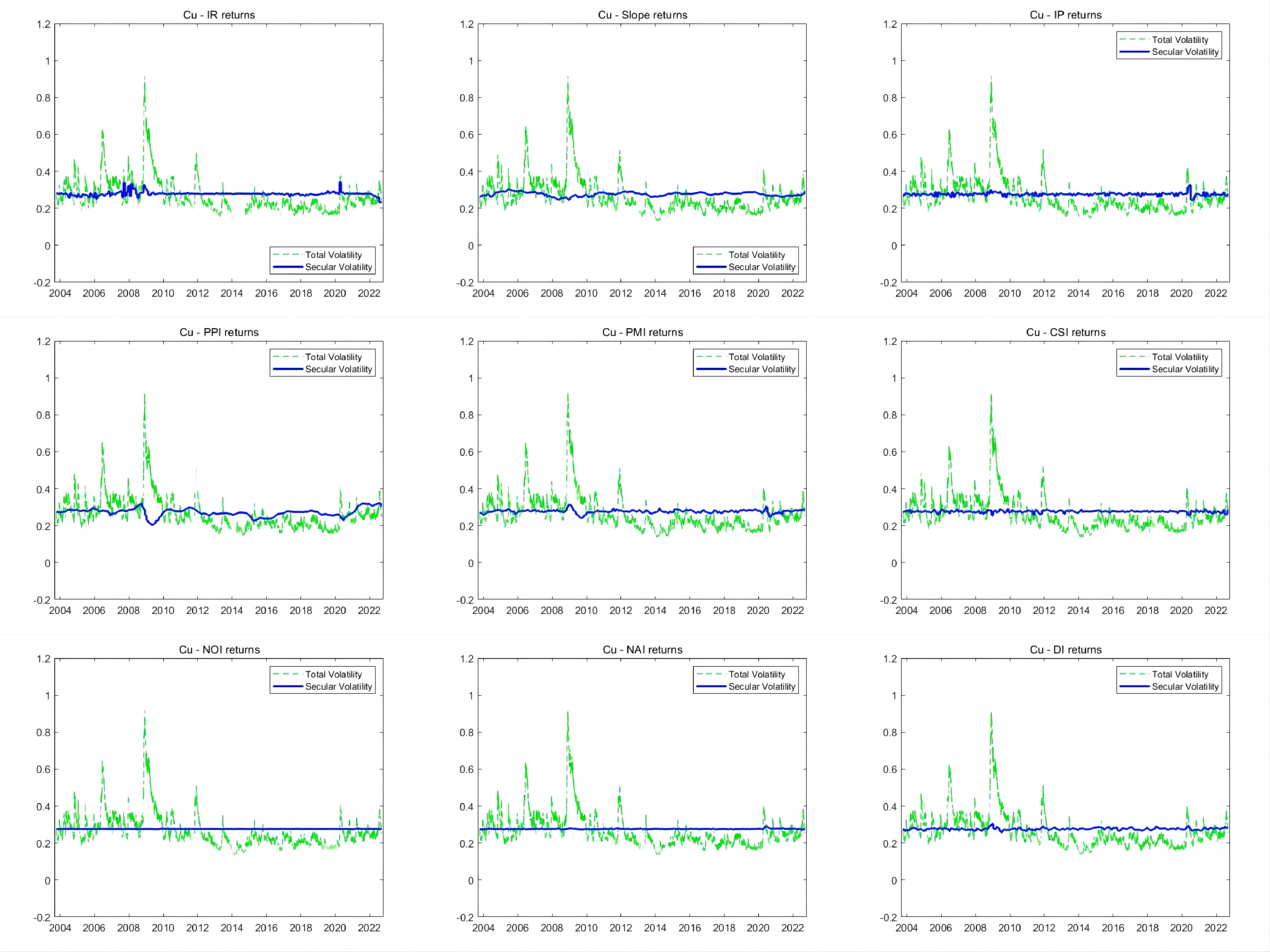}
\end{figure}

In Figure 3 and Figure 4, the estimated conditional volatility and its long-run component of the GARCH-MIDAS model with the return and volatility of macroeconomic variables are presented. The lag length K is 250 for daily variables (IR and Slope), and 12 for monthly variables. The plots illustrate the statistic results in Table 4. The larger absolute value of significant $\theta$ , and larger VR, the more the variable contribute to the conditional variance of Cu return as a long run component in the MIDAS filter. For the models with index levels (return), PPI's model shows significant positive effect to the conditional volatility of copper future returns, especially during the 2008-2010 crisis, it is clear that the long run component decreases quite earlier than the total volatility. IR's return has negative $\theta$ value, and in its figure the long run component sometimes goes opposite to the volatility's direction. This is also true for IP, but its $\theta$ is rather small, so the long-run curve is less volatile.

In Figure 4, the curves of model with volatility are more volatile than those with index levels. Slope has the largest positive $\theta$ and VR value, as can be seen in its plot, the long-run component contributes to the total volatility. PPI's $\theta$ is also significant in the model with volatility and it shows positive effect on the copper volatility. CSI and DI's volatility also have significant positive influence to the total volatility, whereas DI's curve is more volatile but with smaller variation. It means that when the market become volatile, the consumer sentiment index which reflects the consumer's expecting behavior, will have greater impact on the copper future returns more than the US dollar index, which reflects this currency's strength.

\begin{figure}[H]
    \caption{Conditional volatility and its long run component of Cu future returns - Based on Index Volatilities}
    \label{Figure4}
    \includegraphics[width=\textwidth,height=0.5\textheight]{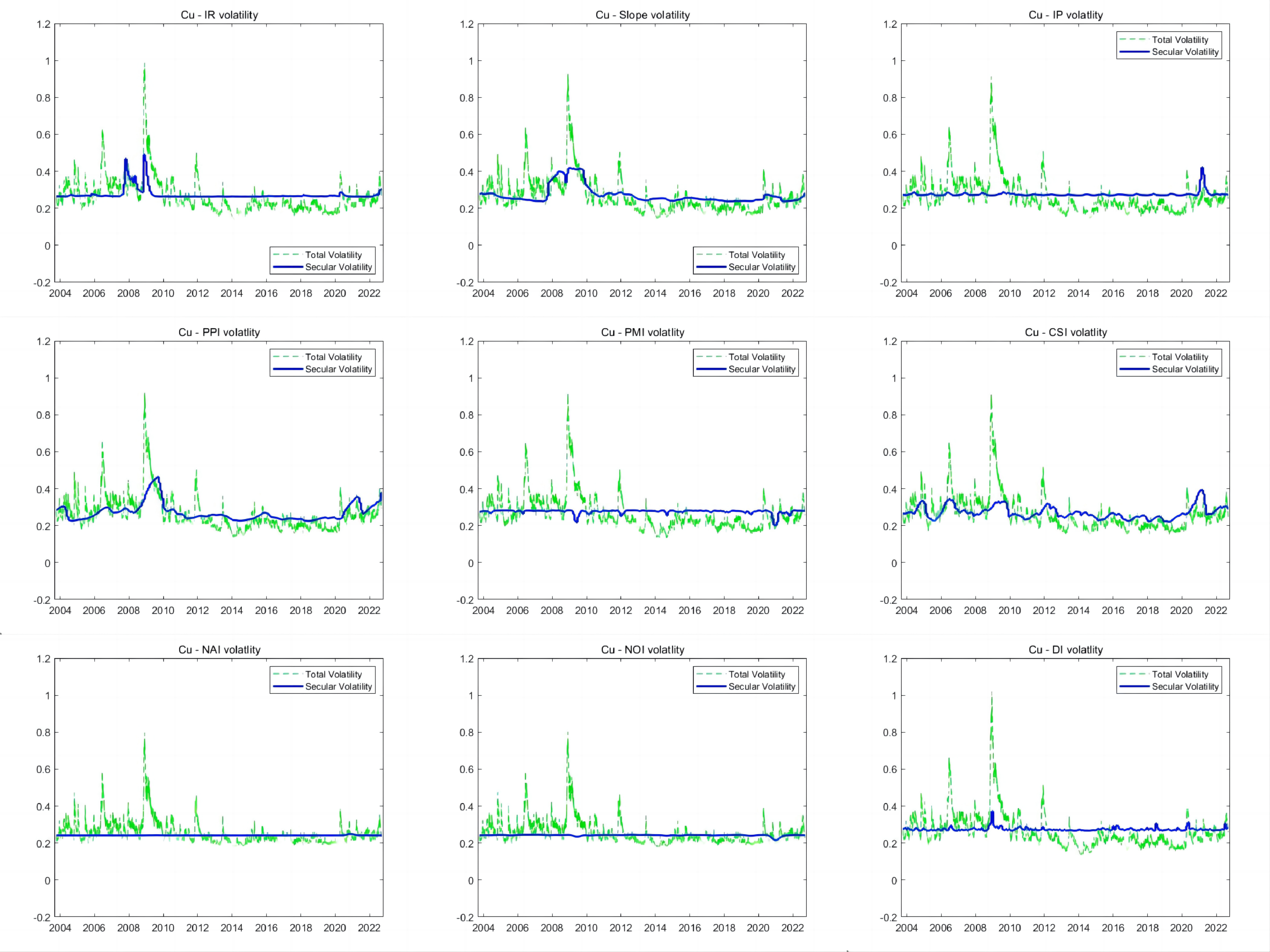}
\end{figure}

\subsection{Bivariate GARCH-MIDAS model}

In the previous sections, we analysed the performance of univariate GARCH-MIDAS model with independent macroeconomic variables. Next, we are going to put them into bivariate models, in pairs of RV+X (realized volatility+macroeconomic variable), $X_k$ level+$X_j$ level, $X_k$ volatility+$X_j$ volatility, and $X_k$ level+$X_j$ volatility.

In Table \ref{Table5}, the estimation results of RV+X model with significant $\theta$ values for both variables are presented. Comparing to the univariate models, some of the previously significant $\theta$s are not significant in the bivariate model. The $\alpha$ and $\beta$ of all models are still significant and have sums larger than 0.9.
{
\begin{table}[H]
    \footnotesize
    \centering
    \caption{Bivariate model of RV with index level/volatility}
    \label{Table5}
    \resizebox{\textwidth}{!}{%
        \begin{tabular}{@{}lllllllllll@{}}
            \toprule
                        & $\alpha$  & $\beta$   & m         & $\theta^{x_1}$ & $\omega_2^{x_1}$ & $\theta^{x_2}$ & $\omega_2^{x_2}$ & BIC      & LLH      & VR(X) \\ \midrule
            Index Level &           &           &           &                &                  &                &                  &          &          &       \\
            RV+IP       & 0.0624*** & 0.8958*** & 0.5358*** & 0.0071***      & 1.0903**         & -0.0298*       & 23.7853**        & 17735.30 & -8829.61 & 39.20 \\
                        & (0.0029)  & (0.0000)  & (0.0000)  & (0.0000)       & (0.0416)         & (0.0853)       & (0.0487)         &          &          &       \\
            RV+PPI      & 0.0571*** & 0.8879*** & 0.2974*** & 0.0089***      & 1.2422***        & 0.3685***      & 1.0001*          & 17717.77 & -8820.84 & 57.98 \\
                        & (0.0008)  & (0.0000)  & (0.0045)  & (0.0000)       & (0.0056)         & (0.0001)       & (0.0729)         &          &          &       \\
            RV+NAI      & 0.0644*** & 0.8788*** & 0.3997*** & 0.0098***      & 1.2215***        & 0.0040**       & 1.0002***        & 17726.78 & -8825.35 & 48.31 \\
                        & (0.0006)  & (0.0000)  & (0.0001)  & (0.0000)       & (0.0022)         & (0.0135)       & (0.0041)         &          &          &       \\
            \midrule
            Volatility  &           &           &           &                &                  &                &                  &          &          &       \\
            RV+IR       & 0.0702*** & 0.8691*** & 0.4618*** & 0.0070***      & 1.3206**         & 0.3476**       & 4.2423           & 17735.52 & -8829.72 & 60.81 \\
                        & (0.0026)  & (0.0000)  & (0.0000)  & (0.0000)       & (0.0271)         & (0.0219)       & (0.1601)         &          &          &       \\
            RV+PPI      & 0.0617*** & 0.9023*** & 0.5883*** & 0.0069***      & 1.0190           & 0.0000*        & 39.3777***       & 17732.71 & -8828.32 & 45.68 \\
                        & (0.0085)  & (0.0000)  & (0.0000)  & (0.0000)       & (0.1402)         & (0.0721)       & (0.0000)         &          &          &       \\
            RV+CSI      & 0.0661*** & 0.8646*** & 0.2629**  & 0.0067***      & 1.2055***        & 0.0001***      & 1.1703***        & 17727.40 & -8825.66 & 44.16 \\
                        & (0.0005)  & (0.0000)  & (0.0277)  & (0.0000)       & (0.0016)         & (0.0063)       & (0.0005)         &          &          &       \\ \bottomrule
        \end{tabular}%
    }
\end{table}
}
\noindent
\textit{Note: Numbers in parenthesis are the p-value of each estimated parameter. ***, **, and * indicate significance at the 1$\%$, 5$\%$, and 10$\%$ levels, respectively. Estimates for $\mu$ and $\gamma$ are omitted. Only bivariate models with significant estimation value of $\theta$ of both variables are shown here.}

For the RV+level model, NAI's existence increases the $\theta$ of RV, but its own $\theta$ is relatively small though significant. PPI and IP's $\theta$'s absolute value both increase slightly, along with the RV's $\theta$, which indicates larger influence to the copper future return. For the RV+volatility models, the PPI and CSI's performance are not that noticeable, which is consistent to their behaviour in the univariate model. The overall performance of RV+X model is generally better than the univariate model of RV, with increment in LLH and VR, and decrement in BIC. In perspective of RV's $\theta$, its value is decreased in the RV+volatility model, and slightly increased in RV+level model.

The bivariate GARCH-MIDAS model with combinations of macroeconomic variable's level and volatility are shown in Table \ref{Table6}, in which we filter and present the models with both significant $\theta$s.

\begin{table}[H]
    \footnotesize
    \centering
    \caption{Bivariate model of index level/volatility of macroeconomic variables}
    \label{Table6}
    \resizebox{\textwidth}{!}{%
        \begin{tabular}{@{}lllllllllll@{}}
            \toprule
                        & $\alpha$  & $\beta$   & \textit{m}         & $\theta^{x_1}$ & $\omega_2^{x_1}$ & $\theta^{x_2}$ & $\omega_2^{x_2}$ & BIC      & LLH      & VR(X) \\ \midrule
            Index Level &           &           &           &                &                  &                &                  &          &          &       \\
            IR+IP       & 0.0401*** & 0.9495*** & 0.9779*** & -0.2004**      & 10.8483***       & -0.0308*       & 30.5712**        & 17732.42 & -8828.17 & 8.58  \\
                        & (0.0028)  & (0.0000)  & (0.0000)  & (0.0184)       & (0.0004)         & (0.0513)       & (0.0212)         &          &          &       \\
            IR+PPI      & 0.0430*** & 0.9427*** & 0.8730*** & -0.2235**      & 9.1837***        & 0.3595**       & 1.7139**         & 17728.86 & -8826.39 & 18.53 \\
                        & (0.0023)  & (0.0000)  & (0.0000)  & (0.0205)       & (0.0033)         & (0.0213)       & (0.0352)         &          &          &       \\
            IP+PPI      & 0.0498*** & 0.9327*** & 0.8938*** & -0.0329**      & 27.3572**        & 0.3426**       & 1.9743*          & 17746.17 & -8835.04 & 15.29 \\
                        & (0.0014)  & (0.0000)  & (0.0000)  & (0.0389)       & (0.0367)         & (0.0135)       & (0.0666)         &          &          &       \\
            PPI+NOI     & 0.0506*** & 0.9320*** & 0.8946*** & 0.3289**       & 2.0116           & -0.0081*       & 26.2914**        & 17746.48 & -8835.20 & 16.40 \\
                        & (0.0017)  & (0.0000)  & (0.0000)  & (0.0259)       & (0.1579)         & (0.0768)       & (0.0327)         &          &          &       \\ \midrule
            Volatility  &           &           &           &                &                  &                &                  &          &          &       \\
            IR+CSI      & 0.0762*** & 0.8690*** & 0.5669*** & 0.5855***      & 1.4667           & 0.0001*        & 1.0000***        & 17735.23 & -8829.57 & 25.24 \\
                        & (0.0061)  & (0.0000)  & (0.0021)  & (0.0000)       & (0.1334)         & (0.0525)       & (0.0031)         &          &          &       \\
            Slope+CSI   & 0.0708*** & 0.8699*** & 0.3990**  & 0.6919***      & 1.7496*          & 0.0001*        & 1.0007***        & 17740.17 & -8832.04 & 46.02 \\
                        & (0.0024)  & (0.0000)  & (0.0145)  & (0.0000)       & (0.0628)         & (0.0567)       & (0.0053)         &          &          &       \\
            Slope+DI    & 0.0512*** & 0.9293*** & 0.5966*** & 0.1253*        & 140.7425         & 0.0027***      & 1.0000           & 17727.72 & -8825.82 & 19.40 \\
                        & (0.0042)  & (0.0000)  & (0.0010)  & (0.0693)       & (0.3708)         & (0.0043)       & (0.1175)         &          &          &       \\
            PPI+NOI     & 0.0609*** & 0.8623*** & 0.7216*** & 0.0009***      & 1.0000***        & 0.0000*        & 1.7344**         & 17766.30 & -8845.11 & 46.44 \\
                        & (0.0043)  & (0.0000)  & (0.0000)  & (0.0000)       & (0.0000)         & (0.0506)       & (0.0248)         &          &          &       \\ \midrule
            Mixed       &           &           &           &                &                  &                &                  &          &          &       \\
            PPI         & 0.0592*** & 0.9254*** & 0.9822*** & 0.3493**       & 1.5475**         & 0.0000**       & 37.1982***       & 17744.85 & -8834.38 & 15.01 \\
                        & (0.0029)  & (0.0000)  & (0.0000)  & (0.0213)       & (0.0288)         & (0.0176)       & (0.0017)         &          &          &       \\ \bottomrule
        \end{tabular}%
    }
\end{table}
\noindent
\textit{Note: Numbers in parenthesis are the p-value of each estimated parameter. ***, **, and * indicate
significance at the 1$\%$, 5$\%$, and 10$\%$ levels, respectively. Estimates for $\mu$ and $\gamma$ are omitted. Only bivariate models with significant estimation value of $\theta$ of both variables are shown here. The third panel named “Mixed” means the independent variables are one macroeconomic variable's level value and its own volatility.}

In consistency with the result of univariate models in Table \ref{Table4}, the variables with significant $\theta$s still appear in the level+level model and vol+vol model correspondingly. For the model with two variable's level value, we do not observe a general rule for their $\theta$'s performance. In the model of IP+PPI, both of their $\theta$'s absolute value increase. In the other three models, one of the $\theta$'s absolute value increases as another decreases. NOI's $\theta$ is not significant in its univariate model, but in the PPI+NOI model it has negative effect to the copper future's return, however, the PPI's $\theta$ is decreased, while PPI's $\omega_2$ becomes insignificant. In the bivariate volatility model, Slope, CSI and PPI are still significant. PPI, CSI and DI's $\omega_2$ are still very close to 1, same to the value in Table \ref{Table4}. For IR+CSI model, IR's $\theta$ become significant and relatively large, while CSI's $\theta$ is still small. However, in Slope+CSI/DI's model, Slope's $\theta$ is lower than that in the univariate model. The PPI+NOI's combination is significant for $\theta$'s existence, but the values are too close to zero so that they have little impact to the copper future return.

Generally, the bivariate model with both macroeconomic variables of significant $\theta$ perform better than univariate model. The value of VR is noticeably larger than in univariate models, in which PPI has the largest VR of 14.46 for its return and Slope has the largest VR of 31.28 with its volatility in perspective. While in bivariate models, these values are improved.

For the models with their own index level and volatility, only PPI is outstanding with both $\theta$s significant. As supplement, NOI's index level and DI's volatility also have significant $\theta$, which is consistent to their results in the univariate model, but their mixed bivariate model fail. PPI's performance is also similar to its univariate model: index level $\theta$ is 0.3340 in Table \ref{Table4}, and volatility $\theta$ is 0.0006. In the PPI mixed model, its level's positive impact increases, while its volatility becomes even less effective to the copper future return.

Overall, PPI contributes to all kinds of models we have built, and its index level have more influence to the copper future return. Therefore, for the bivariate models including level values, PPI is the preferred choice. Slope's volatility is also noticeable, though its $\theta$ is not significant in the model with RV, its performance in univariate and bivariate model with other volatilities is quite well.

\subsection{DCC-MIDAS model}
The dynamic conditional correlation between COMEX copper future return and S\&P 500 index return (both log differenced) is estimated by DCC method. Following Engle's (2002) two step approach, the volatility term we apply the GARCH (1,1) to both return series in step one, then take it into DCC (1,1) estimation. The results are shown in \ref{Table7} below. The DCC parameters \textit{a} and \textit{b} are both significant at 0.01 level, and \textit{b} is close to 1, which indicates noticeable effect from previous dynamic correlation and persistence of change. In the second row, we use PPI as \textit{a} covariate of the long-run Cu-S\&P 500 correlation in the DCC-MIDAS-MV model, where PPI's effect is illustrated in section 5.3. In consistence with previous settings of GARCH-MIDAS models, the weighting scheme is still restricted that we fix $\omega_1=1$. The a and b are the parameters of the short-term correlation and their sum is still close to one. Comparatively, the DCC-MIDAS-PPI result performs better than DCC-GARCH, with lower AIC and BIC, and higher Log Likelihood. Moreover, as PPI and RV both performed well in the univariate and bivariate GARCH-MIDAS models, we further put RV in to the DCC-MIDAS model in the third row of Table 7 and make comparison with the one with PPI. Though the GARCH-MIDAS model with RV always has a larger VR value and smaller BIC value, the DCC-MIDAS with PPI outperforms the one with RV. It further confirms the efficiency of PPI in modelling the Cu future returns' volatility.

\begin{table}[H]
    \footnotesize
    \centering
    \caption{DCC-GARCH (1,1) and DCC-MIDAS estimation results of Cu-S\&P 500}
    \label{Table7}
    \resizebox{\textwidth}{!}{%
        \begin{tabular}{@{}lllllll@{}}
            \toprule
                                & \textit{a}         & \textit{b}         & $\omega_2$ & AIC      & BIC      & LLH      \\ \midrule
            DCC-GARCH           & 0.0155*** & 0.9809*** & -          & 17718.62 & 17737.98 & -8856.31 \\
                                & (0.0000)  & (0.0000)  &            &          &          &          \\
            DCC-MIDAS-PPI-level & 0.0126*** & 0.9864*** & 1.7131***  & 17686.17 & 17699.08 & -8841.09 \\
                                & (0.0006)  & (0.0000)  & (0.0000)   &          &          &          \\
            DCC-MIDAS-RV        & 0.0135*** & 0.9855*** & 1.8309***  & 17782.65 & 17802.02 & -8888.33 \\
                                & (0.0003)  & (0.0000)  & (0.0000)   &          &          &          \\
            \bottomrule
        \end{tabular}%
    }
\end{table}
\noindent
\textit{Note: Numbers in parenthesis are the p-value of each estimated parameter. ***, **, and * indicate
significance at the 1$\%$, 5$\%$, and 10$\%$ levels, respectively. The number of MIDAS lags is 12 for PPI, and 24 for the long run correlation of DCC process.}

\begin{figure}[]
    \caption{20 years of Cu-S\&P 500 correlations vs. Cu/S\&P500 return}
    \label{Figure5}
    \includegraphics[width=\textwidth]{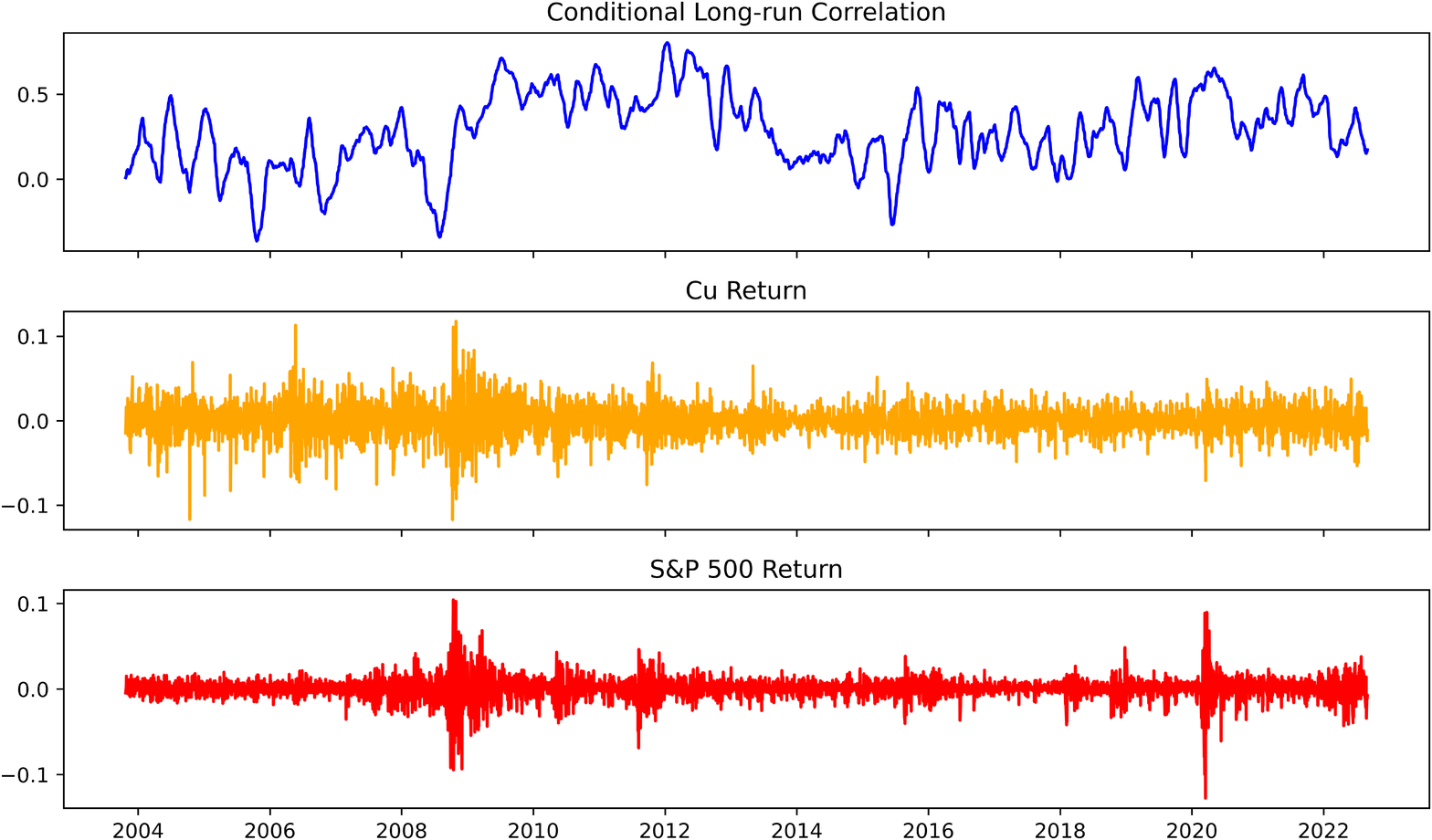}
\end{figure}

The output of dynamic correlations is plotted as in the first panel of Figure \ref{Figure5}, in comparison with the returns of copper future and S\&P 500. The range of the correlation is [-0.3635, 0.8034]. The most negative values close to -0.3 take place in Oct, 2005 and July, 2008, while the return of copper future and S\&P 500 were both volatile, then become mostly positive after the 2008 drop. Although it keeps being mostly positive for the period of 2009 to 2022, there still exists noticeable volatility in the correlation. Comparing to the plot of returns in Figure 3, the April, 2022 increment in the correlation appears in the mean while when the S\&P 500 return suddenly become volatile.

\section{Conclusion}
In this paper, the dynamic relationship of COMEX Copper future return and macroeconomic variables is studied by estimating GARCH-MIDAS models and DCC-MIDAS models. By the results of univariate and bivariate GARCH-MIDAS models, Conrad's (2019) opinion of “Two are better than one” is proved that the bivariate models with both significant $\theta$ values always perform better than the univariate models with the same explanatory variable according to information criteria. Among all the macroeconomic variables, PPI is the most efficient in both its level value and volatility value to influence the copper returns. Next, the DCC-GARCH is used to model the dynamic conditional correlation between Cu and S\&P500, and the DCC-MIDAS-PPI further incorporate PPI to model this relationship, which is proved to be better than simple DCC-GARCH.

Therefore, the level value of IR and IP have reverse effect on volatility of Cu future returns that increment in their value would decrease the volatility of copper future return, showing a countercyclical characteristic, while PPI level has positive relation with Cu future returns. For the volatility of indices, Slope, PPI, CSI, and DI have significant effects on Cu future returns. The realized volatility generated from the Cu future series reasonably has the greatest influence, also for the bivariate models including it and other macroeconomic factors. Moreover, for the dynamic conditional correlation models of Cu future returns and S\&P 500, PPI is also proved to have significant effect in modelling the relationship. Therefore, the influence of PPI on Cu future returns is significant and it is suggested to be used as an explanatory variable for COMEX Copper future.

Some limitations of this research include the restricted weighting scheme of variables and the only choice of GJR-GARCH of the bivariate GARCH-MIDAS model. Future studies are suggested to release the weighting scheme and use standard GARCH (1,1) in GARCH-MIDAS model, and further incorporate three and more explanatory variables.
\clearpage
\nocite{*}
\bibliographystyle{apacite}
\bibliography{myref}
\end{document}